\documentclass[aip,twocolumn,superscriptaddress]{revtex4}
\usepackage{graphicx}
\begin{document}

\title{Quantum key distribution and 1~Gbit/s data encryption over a single fibre}
\author{Patrick Eraerds}
\author{Nino Walenta}
\affiliation{Group of Applied Physics-Optique, University of Geneva, Rue de l'\'Ecole-de-M\'edecine 20, 1205 Geneva, Switzerland}
\author{Matthieu Legr\'e}
\affiliation{idQuantique SA, Chemin de la Marbrerie 3, 1227, Geneva, Switzerland}
\author{Nicolas Gisin}
\author{Hugo Zbinden}
\affiliation{Group of Applied Physics-Optique, University of Geneva, Rue de l'\'Ecole-de-M\'edecine 20, 1205 Geneva, Switzerland}
\date{\today}

\begin{abstract}
We perform quantum key distribution (QKD) in the presence of 4 classical channels in a C-band dense wavelength division multiplexing (DWDM) configuration using a commercial QKD system. The classical channels are used for key distillation and 1~Gbps encrypted communication, rendering the entire system independent from any other communication channel than a single dedicated fibre. We successfully distil secret keys over fibre spans of up to 50~km. The separation between quantum channel and nearest classical channel is only 200~GHz, while the classical channels are all separated by 100~GHz. In addition to that we discuss possible improvements and alternative configurations, for instance whether it is advantageous to choose the quantum channel at 1310~nm or to opt for a pure C-band configuration.
\end{abstract}
\maketitle
\section{Introduction}\label{sec:Intro}
Since the initial proposal of quantum key distribution (QKD) in 1984 \cite{BB84} and its first experimental demonstration \cite{Bennett92}, major progresses in long-distance, fibre-based point-to-point QKD have been achieved (for an overview over current state-of-the-art implementations, see \cite{NJPFocusOnQKD}).

Until recently, one of the specifics of QKD systems was the need for a dedicated dark optical fibre, exclusively reserved for the quantum channel (single photon level). Signals of classical strength, used to perform key distillation and encrypted communication between the end users, were sent through a second fibre to not compromise the weak quantum signal.

The next consequential step towards larger availability of QKD links is to look at the compatibility of QKD with existing fibre infrastructure. Common public DWDM (dense wavelength division multiplexing) networks multiplex up to 50 different wavelength channels on a single fibre.
If the quantum channel is launched into a fibre accompanied with other classical signals, several effects like channel crosstalk, Raman scattering, four-wave mixing or amplified spontaneous emission (in case of amplification of the classical channels) can severely degrade QKD system operation or worse, can prevent it at all.

First investigations in this direction were conducted by Townsend in the late nineties \cite{Townsend}. The impact of a single classical C-band channel, wavelength multiplexed with a quantum channel at 1310~nm, was analyzed. Later, in 2005, Lee and Wellbrock demonstrated quantum key distribution, placing both the quantum channel and one classical channel into the C-band with a separation of down to 400~GHz equivalent to 3.2~nm \cite{Lee}. We note that the classical channel was neither linked to QKD system operation nor used for encrypted communication. More recent works  \cite{Telcordia2009,Telcordia2007} investigate different impairment sources on a more general level, including effects which occur when more than one classical channel is present, e.g. four-wave mixing.

Apart from the long term goal of QKD operation on public DWDM networks, another frequently encountered network topology could push forward QKD availability on a short term scale. In order to accommodate future growth, telecom companies have spent the last few years installing point-to-point dedicated fibres. These fibres can also be used in the standard configuration of QKD using a dark fibre for the quantum channel and another for the encrypted communication. However, for reasons of availability and fibre leasing costs the operation on only one fibre is highly desirable.

In this paper we investigate exactly this situation where in total only one dedicated fibre is available and an encrypted link, based on QKD, should be established between its endpoints. This objective thus necessitates the wavelength multiplexing of all system relevant channels, i.e. key distillation- and encrypted communication channels as well as the quantum channel on a single fibre. If one is not obliged to operate in a two fibre configuration, then in particular QKD systems which require a classical clock signal to synchronize the separate devices would benefit from higher robustness against fibre drifts. Furthermore, the configuration investigated here surely bears the advantage of having perfect information on the classical channels, while they are difficult to assess in the public network configuration. Therefore a reliable performance characterisation of the entire system is obtainable.

In our experiment we use a standard 8 channel C-band DWDM with 100~GHz (corresponding to 0.8~nm) spacing. We simultaneously multiplex 4 classical channels (one bidirectional channel for distillation and encrypted 1~Gbps communication, respectively) with a quantum channel, separated from the nearest classical channel by only 200~GHz.

The paper is organized as follows: In Sec.~\ref{sec:Impairment} we discuss the different impairment sources relevant for our realization. Section~\ref{sec:Experiment} describes the QKD setup and presents the experimental results followed by a discussion and outlook in Sec.~\ref{sec:Discussion}. Section~\ref{sec:Conclusions} contains our conclusions.

\section{Impairment sources}\label{sec:Impairment}
\subsection{Raman scattering}\label{sec:Raman}

\begin{figure}

	\centering	
	\includegraphics[width=1\linewidth]{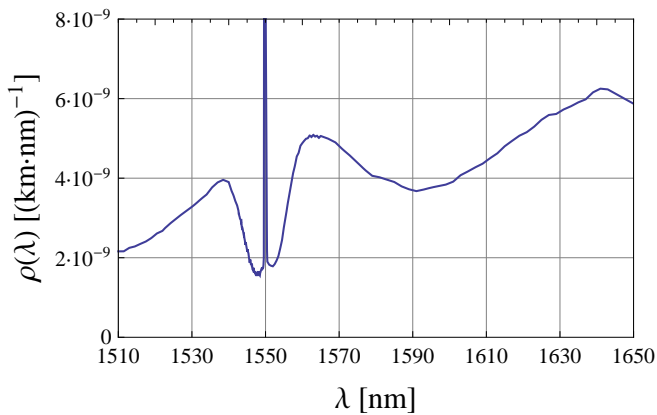}
	\caption{\label{fig:RamanSpectra} Measured effective Raman cross-section $\rho\left(\lambda\right)$ (per km fibre length and nm bandwidth) for a pump laser wavelength centred at 1550~nm in a standard single mode fibre at room temperature.}
\end{figure}
\begin{figure}
	\centering
	\includegraphics[width=1\linewidth]{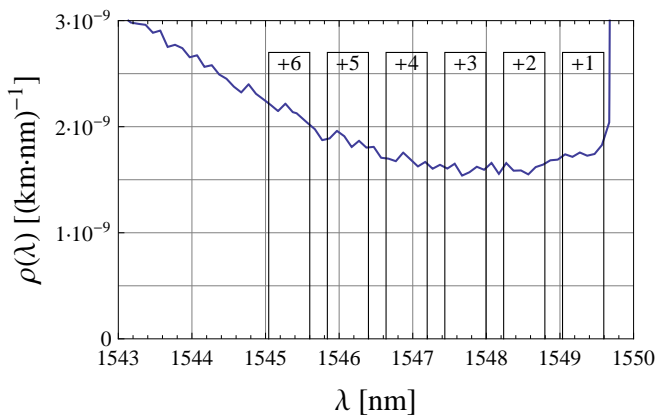}

\caption{\label{fig:RamanSpectra} Zoom on Anti-Stokes dip of the Raman spectrum. In channels +2 and +3 the minimal amount of Raman scatter is found.}
\end{figure}

Due to photon-phonon interaction, photons can change their wavelength and thus compromise other channels. Depending on whether a phonon gets excited or de-exited, photons at wavelengths above (Stokes) and below (anti-Stokes) the initial wavelength are generated. Scattering off acoustic phonons (Brillouin scattering) is not critical, since the maximal frequency shift of the scattered photons is small (10~GHz, in backward direction) and therefore cannot reach adjacent channels on a 100~GHz grid. By contrast, scattering off optical phonons (Raman scattering) can lead to significant frequency shifts covering the entire C-band~\footnote{Assuming the initial pump frequency to lie somewhere in the C-band.}, having an intensity maximum at a shift of about 13~THz (corresponding to a wavelength shift of 100~nm at 1550~nm). Unlike acoustic phonons, the more or less flat dispersion relation of optical phonons causes frequency shifts independent of the scatter direction. This means that in co- as well as in counter propagating direction (with respect to the exciting signal) a broad spectrum of photons is present.

We measured the Raman scatter generated by a 50 km standard single mode fibre and extracted an effective Raman scattering cross-section $\rho(\lambda)$, shown in Fig.~\ref{fig:RamanSpectra}. It is normalized with respect to spectral bandwidth and fibre length and accounts for the fibre caption ratio of the scattered light. In return, by means of $\rho(\lambda)$ and allowing for fibre attenuation we can calculate the Raman scatter power emerging from the input $P_{ram,b}$ (backward Raman scattering) and output $P_{ram,f}$ (forward Raman scattering) of a fibre of  arbitrary length $L$. Assuming a certain filter pass-band $[\lambda,\lambda+\Delta\lambda]$ and approximating the spectral integration via
\begin{center}
\begin{equation}
	\int^{\lambda+\Delta\lambda}_{\lambda}\rho(\lambda')d\lambda' \approx \rho(\lambda)\cdot \Delta\lambda
	\label{Ispec}
\end{equation}
\end{center}
we obtain (see Appendix~A)
\begin{equation}
	P_{ram,f} = P_{out}\cdot L\cdot \rho(\lambda)\cdot \Delta\lambda \\\\
	\label{eq:RamanFW}
\end{equation}
\begin{equation}
	P_{ram,b} = P_{out}\cdot \frac{sinh(\alpha\cdot L)}{\alpha}\cdot \rho(\lambda)\cdot \Delta\lambda
	\label{eq:RamanBW}
\end{equation}
where $P_{out}$ is the power of the exciting laser at the fibre output in [W], $\alpha$ the fibre attenuation coefficient [km$^{-1}$] and $L$ the fibre length [km]. $P_{out}$ can be written in terms of the input power via $P_{out}=P_{in}\cdot e^{-\alpha\cdot L}$. The impact of each of the Raman contributions, represented by the detection probability per ns detector gate is depicted in Fig.~\ref{fig:AllNoise}.

Note that we assume equal attenuation for initial and scattered wavelength, which is reasonable for our total wavelength span of 4~nm (see Sec.~\ref{sec:Experiment}).

\subsection{Channel crosstalk}\label{sec:CrossTalk}
The relative strength of the classical channels requires a large DWDM isolation with respect to the quantum channel. As a reasonable benchmark for a sufficient isolation we propose the detector dark count probability. To calculate a typical value we need to consider the receiver sensitivity of the transceiver modules used for the classical communication (see Sec.~\ref{sec:Experiment}). It conditions the necessary optical power to ensure error free detection.

In our particular case the sensitivity which guarantees a bit error rate $BER<10^{-12}$ is equal to -28~dBm (Finisar FWLF-1631-xx). This power corresponds to approximately $1.2\cdot10^4$ photons per ns. In order to attenuate this photon number so that the detection probability per ns gate is of order of the dark count probability ($5\cdot10^{-6}$~ns$^{-1}$) an isolation of about 80~dB is needed. Here we assume a detector efficiency of $\eta=0.07$ and internal components loss of 2.65~dB on Bob's side.

\begin{figure}[tbp]
 \centering
 \includegraphics[width=\linewidth]{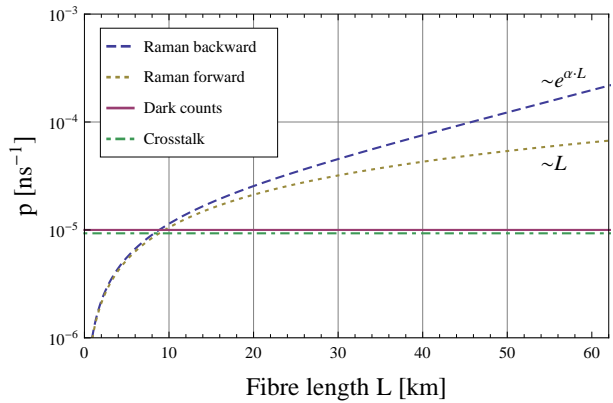}
 \caption{\label{fig:AllNoise} Different contributions to the total noise count probability per ns detector gate assuming our system parameters (DWDM channel isolation=82~dB, fibre loss $\alpha_{dB}=0.21$~dB/km, 4 classical channels each with a power of -28~dBm at the receiver, $\eta=0.07$, internal components loss=2.65~dB).}
\end{figure}

Our standard 8-channel DWDM provides an isolation of 82~dB between non-adjacent channels which is just sufficient to match the before mentioned criteria, see Fig.~\ref{fig:AllNoise}. In the case of insufficient isolation, additional filters can further improve the isolation, however at the expense of additional insertion loss in the quantum channel. In particular, considering Raman scattering we find that crosstalk is not a limiting factor for long fibre lengths. 
 
Finally, we note that a sufficient isolation of the co-propagating quantum and classical channel entails that crosstalk from Rayleigh backscatter in a counter-propagating configuration can be neglected.

\subsection{Four-wave mixing}\label{sec:FWM}
Four-wave mixing (FWM) is mediated by the third order susceptibility $\chi^{(3)}$ and describes the generation of additional photon frequencies, different from those present in the initial fields. In contrast to Raman scattering no energy is transferred to or taken from the fibre, i.e. no phonon excitation or de-excitation takes place. Most harmful for our setup would be the degenerate case where two exciting frequencies $f_1$, $f_2$ (assuming $f_1>f_2$) generate side band frequencies $f_+=f_1+(f_1-f_2)$ and $f_-=f_2-(f_1-f_2)$. If the channel separation is not properly chosen, $f_{+/-}$ may coincide with the quantum channel pass-band. The generation efficiency depends on the phase-matching condition, as well as on the relative polarisation and propagation direction of the involved field frequency components. It is particularly easy to fulfil around the zero dispersion wavelength, where it can corrupt even classical communication \cite{Forghieri}. In Sec.~\ref{sec:Experiment} we present a channel configuration which prevents efficient FWM generation, independent of the fibre type (SSMF, DSF or NZDSF).

In addition to the stimulated case described before, it is also important to assess the noise contribution from spontaneous FWM. Spontaneous FWM allows the creation of signal and idler frequencies $f_s,f_i$ from each pump frequency $f_p$, satisfying energy conservation via $2f_p=f_s+f_i$. 
The efficient generation again depends on the phase-matching condition. Around the zero dispersion wavelength the generated spectrum can be rather broad, superposing the spectrum generated by Raman scattering \cite{Agrawal}. Following \cite{Agrawal} we calculate that even in our most demanding configuration the $\gamma P_0 L$ product is very small ($=0.002$, for considerable contributions at least $\gamma P_0 L$ of about 0.1 is needed). This indicates that even when we were operating around the zero dispersion wavelength, spontaneous FWM can be neglected with respect to Raman scattering.

\section{Experiment}\label{sec:Experiment}

\subsection{Setup}\label{sec:Setup}
\begin{figure*}[btp]
 \includegraphics[width=\linewidth]{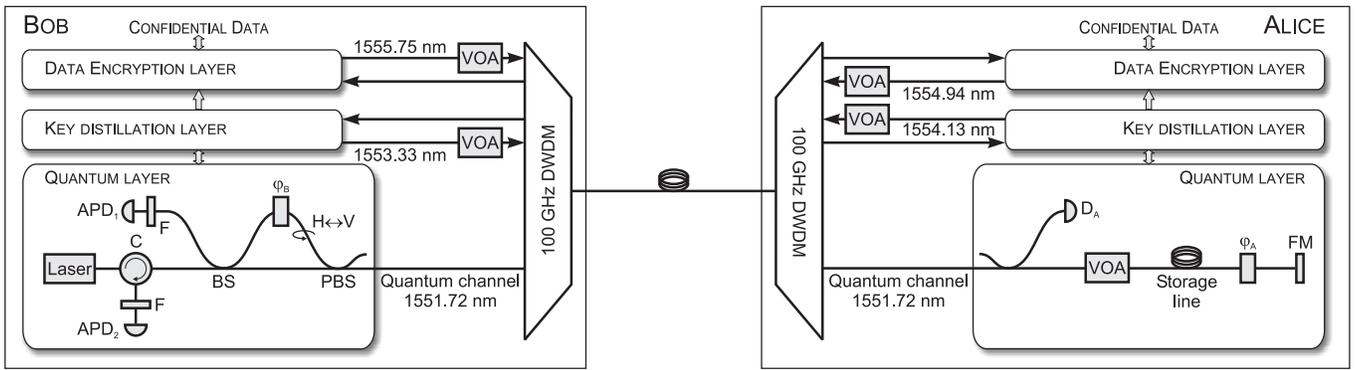}
 \caption{\label{fig:OverallSetup} Setup scheme. APD:~Avalanche photo diode, BS:~Beam splitter, C:~Circulator, D:~Photo diode, F:~Spectral filter (optional), FM:~Faraday mirror, PBS:~Polarizing beam splitter, VOA:~Variable optical attenuator, $\phi$:~Phase modulator.}
\end{figure*}

For the experiments we adopt a commercial QKD system (Cerberis from idQuantique \cite{idQuantique}). As outlined in Fig.~\ref{fig:OverallSetup}, this solution combines a QKD server for secure point-to-point key distribution, and Layer~2 encryption units to encode and decode messages with the key provided by the quantum server for complete secure bidirectional communication between two distant partners, Alice and Bob.

The QKD layer is based on a ``plug \& play'' phase encoding quantum key distribution system where all optical and mechanical fluctuations are automatically and passively compensated \cite{StuckiPlugAndPlay2002}. Bob generates a sequence of optical pulses with a frequency of $f_{rep}=5$~MHz. It propagates through his unbalanced Mach-Zehnder interferometer such that each pulse is split into two orthogonally polarized pulses which are separated by the interferometer imbalance. The sequence length is chosen to match twice the length of the storage line of $L_{s}\approx 10$~km at Alice's in order to avoid compromising Rayleigh backscatter. At Alice's, the major proportion of photons per pulse is used to trigger the classical detector $D_{A}$ in order to synchronize her device with Bob's. The remaining proportion is reflected at the Faraday mirror (FM), phase modulated by $\phi_{A}$ in accordance to Alice's choice of bit value and encoding base, attenuated by the variable optical attenuator (VOA) to $\mu$ photons per pulse and returned to Bob through the same fibre link. Due to the Faraday rotation, each pulse propagates along the contrary interferometer arm as before and interferes at the beam splitter (BS) in accordance to the phase difference between $\phi_{A}$ and Bob's base choice $\phi_{B}$. All internal losses of Bob's optical components sum up to $t_{B}=2.65~dB$ (excluding DWDMs and optional filters).

The signals are detected by InGaAs avalanche photo diodes (APDs) operated in Geiger mode. The APDs are temperature stabilized at 220~K, gated using 1.5~ns long gates and with a dead time of $\tau_{dead}=10~\mu$s applied after each detection to reduce the afterpulse probability to $p_{ap}\approx 0.008$. Their detection efficiencies are $\eta\approx0.07$ at a dark count probability of $p_{dc}\approx5\cdot10^{-6}$~ns$^{-1}$. After key sifting, optionally via the sifting protocols BB84 or SARG \cite{SARG2004}, followed by fully implemented error correction using the CASCADE algorithm \cite{CASCADE1994} and privacy amplification using hashing functions based on Toeplitz matrices \cite{WegmanCarter1981}, Alice and Bob remain with shared secret keys. The integrity of the public distillation communication is ensured by a Wegman-Carter-type authentication scheme based on universal hashing functions \cite{Carter1979}.

The pair of Ethernet encryptors is periodically updated with the secret keys to establish a permanent AES-256 encrypted 1~Gbps~data link between Alice and Bob. The data to be encrypted are continuously provided by two 1~Gbps streams of random bits from a network test system (EXFO PacketBlazer FTB-8510). We note that typically the key refresh rate is once per minute which requires a secret key rate of at least 8.6~bps. In order to guarantee continuous operation the key refresh rate is temporarily reduced if the secret key rate drops below that limit. 

All in all, to completely operate the Cerberis system, four classical communication channels have to be set up between Alice and Bob in addition to the quantum channel. The bidirectional communication for distillation, i.e. key sifting, error correction and privacy amplification, demands two authenticated channels, one from Alice to Bob and one from Bob to Alice. Similarly, two channels are required for the bidirectional encrypted data transmission between the encryptors.
All classical communication channels are implemented using standard optical 2.67~Gbps DWDM SFP transceivers (Finisar FWLF-1631-xx).

We multiplex the quantum channel along with the four classical channels using off-the-shelf 100~GHz DWDM modules (OptiWorks). The modules possess an insertion loss of 1.95~dB and an isolation of 59~dB (82~dB) for adjacent (non-adjacent) channels. The implemented channel configuration is shown in Fig.~\ref{fig:OverallSetup}. For the quantum channel we choose a wavelength of 1551.72~nm on the ITU~C-band grid. We take advantage of 10~\% less Raman noise on the anti-Stokes side of the Raman spectrum at ambient temperature (see Fig.~\ref{fig:RamanSpectra}) by placing all classical channels at higher wavelengths. To benefit from both, the considerably higher DWDM channel isolation for non-adjacent channels as well as from lower Raman noise we omit the adjacent channel and set up the quantum channel 200~GHz (1.6~nm) apart from the nearest classical channel. We minimize the direct impairment due to FWM by choosing the frequency difference between two co-propagating channels such that no FWM frequency product is generated within the quantum channel pass-band (see Sec.~\ref{sec:FWM}).

The discussion of impairment sources has shown that in general the amount of noise impinging on the detectors increases with the total power present in the fibre. Hence, we reduce the power of the classical channels to the overall transmission losses using variable optical attenuators (VOA), such that the corresponding power at the receiver's end just matches the receiver sensitivity of -28~dBm. This corresponds to $P_{out}=-26.05$~dBm in Eq.~\ref{eq:RamanFW},~\ref{eq:RamanBW} due to the insertion losses of our DWDM modules.

With the aim to further minimize the amount of Raman noise, we optionally add phase-shifted fibre Bragg grating filters (F) from aos~\cite{aos} centred around the quantum channel wavelength in front of each APD. Their spectral bandwidth of 45~pm (fwhm) and extinction ratio of 14~dB entails a 85~\% rejection of noise photons, outweighing the additional attenuation of 2~dB due to insertion loss. The filters are actively and independently temperature stabilized using standard temperature controllers, mainly to permit fine adjustment of their transmission bandwidth. A straightforward configuration with only one filter inserted between the PBS and the DWDM was abandoned because of back reflections of the quantum channel laser which completely saturated the APDs.
\begin{table*}
\footnotesize
\centering
\begin{tabular}{@{}lccccccc}
\hline
			Fibre length & 1~km & 5~km & 10~km & 25~km & 35~km & 41~km & 50~km\\
\hline
			\textbf{Without filters}& & & & & & & \\
			$R_{sec}$ [bps] BB84/SARG & 2829/- & 2047/- & 1524/- & 134/511 & 4.3/72 & -/2.0 & \\
			$QBER$ [\%] BB84/SARG & 0.57/- & 0.72/- & 1.18/- & 4.53/2.12 & 8.60/4.77 & -/7.48 & \\
\hline			
			\textbf{With filters}& & & & & & & \\
			$R_{sec}$ [bps] BB84/SARG & & & & 251/347 & 25/128 & 7.5/43 & 0/11\\
			$QBER$ [\%] BB84/SARG &  & & & 1.6/1.7 & 3.6/2.5 & 6.7/3.7 & 34.5/5.4\\
\hline			
		\end{tabular}
		\caption{\label{tab:ResultTable}The secret key rate $R_{sec}$ and $QBER$ values from Fig.~\ref{fig:OverallResults}, which we obtained experimentally using BB84 and SARG, without and with the spectral filters (F).}
\end{table*}

\subsection{Results}\label{sec:Results}
\begin{figure}[b]
	\centering
	\includegraphics[width=1.0\linewidth]{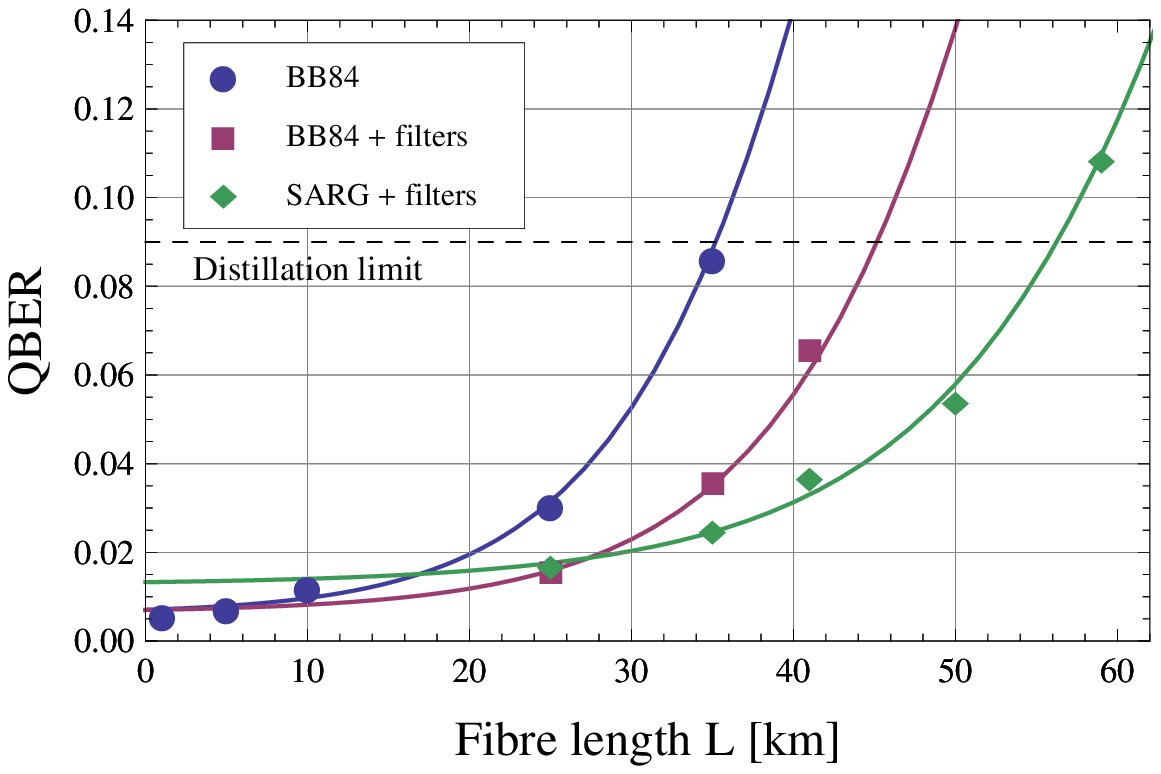}\\
	\includegraphics[width=1.0\linewidth]{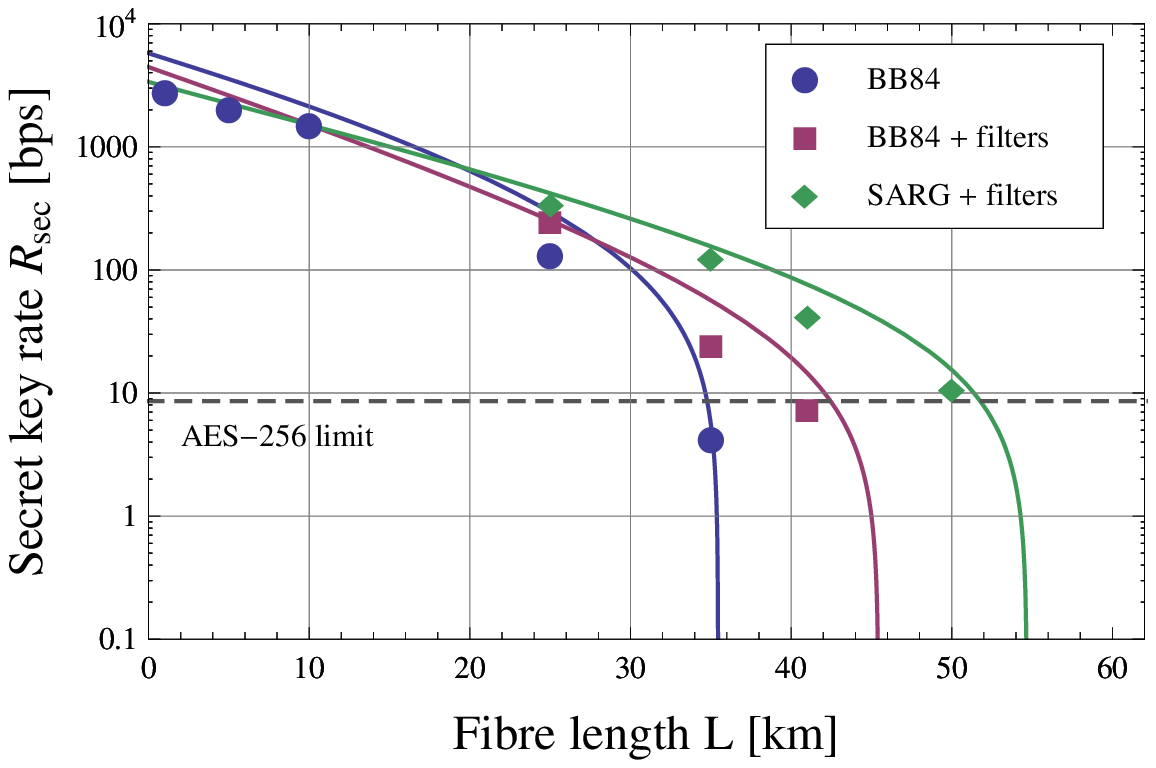}
	\caption{\label{fig:OverallResults}Performance of the QKD based encryption system in terms of $QBER$ (top) and secret key rate provided to the encryptors (bottom) in dependence of the fibre length. Symbols denote our experimental results, solid lines our calculations. Additional filtering increases the maximum fibre length to 41~km using BB84 key sifting and 50~km using SARG.}
\end{figure}

We characterize the system performance for different fibre lengths by measuring the quantum bit error rate $QBER$ and the secret key rate $R_{sec}$. The $QBER$, i.e. the number of erroneous detections over the total number of detections, can be approximated by
\begin{equation}
QBER = QBER_{opt}+QBER_{det}+QBER_{wdm}
\label{eq:DefQBER}
\end{equation}
(for more details see \ref{sec:AppendixQBER}). The optical share $QBER_{opt}$ is determined by the interference visibility entailed by the quality of the optical components and their alignment. Its typical value was 0.3~\% (0.6~\%) using BB84 (SARG). $QBER_{det}$ depends on the characteristics of Bob's single photon detectors and includes errors due to detector dark counts of around $5\cdot10^{-6}$ per ns as well as afterpulses.
$QBER_{wdm}$ summarizes all additional errors from noise due to wavelength-division multiplexing with classical channels, i.e. channel crosstalk and Raman scatter (see Fig.~\ref{fig:AllNoise}).

The secret key rate, i.e. the net rate of secret key bits provided to the encryptors to cipher data communication between Alice and Bob, is given by~\cite{RibordyPlugAndPlay2000}
\begin{equation}
R_{sec} = R_{sift}\left(1-r_{ec}\right)\left(1-r_{pa}\right).
\label{eq:DefSecretRate}
\end{equation}
Here, $R_{sift}$ is the detection rate after sifting (Eq.~\ref{eq:siftRate}), and $r_{ec}$ and $r_{pa}$ are the fractions of bits used for error correction and privacy amplification. Both, $r_{ec}$ and $r_{pa}$, increase non-linearly with the $QBER$.

Our performance results in terms of $QBER$ (estimated by the CASCADE error correction protocol) and net rate of secret keys are plotted in Fig.~\ref{fig:OverallResults} and listed in Tab.~\ref{tab:ResultTable}. The solid lines indicate our calculations which make use of the formulas given in \ref{sec:AppendixQBER}. We note that we do not account for the time needed for key distillation and fibre length measurements, which gets the more significant the higher the key rates. Hence, in general we overestimate the secret key rate especially for short fibre lengths.
The dashed lines in Fig.~\ref{fig:OverallResults} indicate the maximum $QBER$ of 9~\% below which the system can distil a secret key, and the minimum secret key rate of 8.6~bps required for AES encryption with 256~bit keys which are updated once a minute, respectively.

Without the optional spectral filters (F) we obtain a secret key rate which remains well above 1000~bps up to a fibre length of 10~km using BB84 key sifting. Inserting the optional spectral filters in front of the APDs does not only increase the secret key rate from 4.3~bps to 25~bps for a fibre length of 35~km, it also increases the maximal distance to 41~km at which we obtain 7.5~bps. We achieve a further increase in the secret key rate and maximum distance if we use the SARG key sifting protocol instead. Here, the average secret key rate is 128~bps for 35~km and 11~bps for 50~km fibre length. We emphasize that the SARG protocol equally guarantees the security of the key material. While for BB84 the optimum mean photon number $\mu$ of the quantum pulses depends on the fibre transmission $t$ according to $\mu_{BB84}=t$, the SARG protocol allows to benefit from a higher mean photon number $\mu_{SARG}=2\cdot\sqrt{t}$ \cite{BranciardSARG}.

Concerning the stability of the setup we verified constant detection and secret key rates over a period as long as five days in the configuration without the additional filters (F). Having added the filters we still observe a constant detection rate in one detector which confirms that a sufficient stabilisation of the filter transmission spectra can be achieved using standard temperature control. However, after a few hours the detection rate in the second detector tends to decrease due to a drift of the transmission spectra of the corresponding filter, most likely caused by a filter fabrication flaw.
\section{Discussion and outlook}\label{sec:Discussion}
In Fig.~\ref{fig:OverallResults} (left) we compare the $QBER$ values obtained experimentally with theoretical calculations which take all discussed noise sources into account. It reproduces very well the measurement results giving us confidence that we have successfully identified the dominant impairment sources present in our implementation. Based on this we discuss some alternative configurations in the following paragraphs.

\begin{figure}
 \centering
 \includegraphics[width=\linewidth]{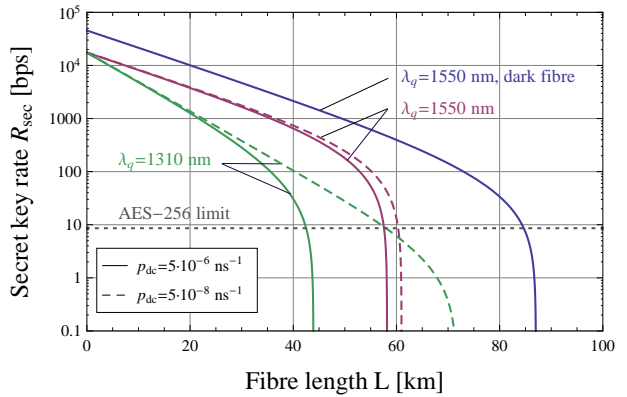}
 \caption{\label{fig:1550vs1310}Comparison between 1550~nm and 1310~nm quantum channel wavelength (SARG, with filters). $p_{dc}$ denotes detector dark count probability. The assumed fibre attenuation is $\alpha_{1550}=0.21$~dB/km and $\alpha_{1310}=0.35$~dB/km. The calculations for a dark fibre configuration (without DWDMs and filters) is also shown for comparison.}
\end{figure}

Firstly, we address the question whether or not it might be advantageous to place the quantum channel in the O-band around 1310~nm while keeping the classical communication channels in the C-band around 1550~nm (for an O-band implementation see \cite{Nist}). The maximal reach of the 1550~nm solution is ultimately limited by the Raman noise (see Fig.~\ref{fig:AllNoise}). Calculating the mean phonon occupation numbers we find that the Raman noise at 1310~nm is about 4000 times weaker as at 1550~nm. We calculate two scenarios: firstly, we take the dark count probability of the detectors used in our experiment (straight lines, $p_{dc}=5\cdot 10^{-6}$~ns$^{-1}$, $\eta = 0.07$) and secondly, we assume a very small detector dark count probability (dashed lines, $p_{dc}=5\cdot10^{-8}$~ns$^{-1}$, $\eta = 0.07$). In addition to that we suppose a better channel isolation in the 1310~nm case of 100~dB while it is at 82~dB in the 1550~nm case (like in our experiment). The results are shown in Fig.~\ref{fig:1550vs1310}. For all curves we neglected the influence of detector dead time, the system specific duty cycle and the reduced efficiency of the error correction protocol (see \ref{sec:AppendixQBER}). As expected, we find that the lower dark count rate dramatically improves the 1310~nm curve, whereas it has rather minor impact on the 1550~nm. However, we see that if high key rates are desired, the 1310~nm solution cannot keep up with the 1550~nm one due to the higher fibre attenuation. Only in an extreme case where lower key rates are acceptable, the 1310~nm solution can reach a larger distance, provided detectors with very low dark count probability are used.

Secondly, we want to estimate the implications of higher transmission rates in the encrypted channels.
As described before, we minimize the total power present in the fibre by adapting the laser power of the SFP modules to their receiver sensitivity of -28~dBm. Modules designated for higher transmission rates currently have lower sensitivity. For example, the 10~Gbps transceiver module Finisar FTRX-1811-3 is specified with a receiver sensitivity of -23~dBm. Using two of these modules for the encrypted link instead of the 1~Gbps modules we used would consequently increase the total classical power by 3.2~dB, and hence the detected noise. Taking this into account but keeping all other parameters unchanged, we estimate for distances up to 40~km no significant degradation of the secret key rate. However, the maximum distance at which a key rate of 8.6~bps can be achieved, reduces by 4-5~km, depending on the sifting protocol.

Next, we take a look on possible measures which could improve the performance of the current setup. One possibility is the reduction of the total classical channel power. This could be achieved by amplification of the classical signals in front of the receivers, or by prospective SFP modules with better receiver sensitivity. While a solution with amplifiers is cost-intensive, an improvement of the receiver sensitivity of more than 3~dB is unlikely in near future. One could also assume that narrower spectral or temporal filtering of the quantum channel could further reduce the impact of Raman noise. However, we observe that there is no more room for improvements here. On one hand the transmission width of 45~pm (corresponding to 5.6~GHz) of our additional filter is already the limit for the spectral width of our sub-nanosecond quantum signals. On the other hand, we can not further reduce the temporal gate width without clipping the pulses and, hence, introducing additional losses. Since the pulse duration of the quantum signals is related to the inverse of its spectral bandwidth, further narrower temporal filtering would entail broader spectral filtering and vice versa.

Finally, we would like to give an outlook on prospective DWDM implementations with next-generation QKD systems based on the Differential-phase shift protocol \cite{DPSYamamoto} or the Coherent-one-way protocol (COW, \cite{COWStucki}). These systems largely benefit from high speed electronics and a better key generation efficiency due to their improved tolerance to photon number splitting attacks. As an illustration we take a look at the COW prototype as presented in \cite{COW250km}, which uses a QKD encoding frequency of 312.5~MHz and a mean photon number of $\mu_{COW}=0.5$~photons per pulse. Assuming the same parameters as used for the calculations with the additional filters in Fig.~\ref{fig:OverallResults} we find an increase in the maximum link distance to 70~km and a secret key rate of $>$~10,000~bps for fibre lengths up to 43~km.

\section{Conclusions}\label{sec:Conclusions}
We demonstrate that a QKD based encryption system can be efficiently operated on a single dedicated fibre of up to 50~km length. All four classical channels necessary to establish the encrypted link are multiplexed along with the quantum signal in a 100~GHz DWDM configuration, rendering the system independent of any additional network connection.
We also show that a combined O-band/C-band solution (quantum channel at 1310~nm) cannot improve the performance.

We find that with respect to the conventional dark fibre configuration, requiring two independent fibres, comparable secret key rates can be obtained, e.g. up to 25~km the decrease of the secret key rate is less than 50~\%. We conclude that with only moderate additional efforts a commercial QKD system can be upgraded to network topologies where only one dedicated fibre is available at a time.

\section{Acknowledgements}
We gratefully acknowledge the helpful support by Patrick Trinkler from idQuantique. This project was financially supported by Swiss NCCR ``Quantum Photonics'' and the ERC-AG QORE.

\section{Appendix}
\subsection{Derivation of Raman Scatter power formulas}\label{App1}
The Raman Scatter power $dP_{ram}$ at wavelength $\lambda$ from a fibre element of length $dx$ at position $x$ when a power $P_{in}$ is launched into a fibre is
\begin{equation}
	dP_{ram}(\lambda,x)= P_{in}\cdot e^{-\alpha\cdot x}\cdot\rho(\lambda)\cdot\Delta\lambda\cdot dx
\end{equation}
$\rho(\lambda)$ accounts already for the fibre caption ratio and we used the same approximation for the spectral integral (see Eq.~\ref{Ispec}). The scatter from a single fibre element is almost isotropic. Now we have to account for the fibre attenuation (fibre length $L$) when the scatter propagates to the fibre output (forward scatter) and back to the fibre input (backward scatter)\\
a) forward :
\begin{equation}
	dP_{ram,f} = dP_{ram}(\lambda,x)\cdot e^{-\alpha\cdot(L-x)}
\end{equation}
integrating over whole fibre 
\begin{equation}
	\Rightarrow P_{ram,f}=P_{in}\cdot e^{-\alpha\cdot L}\cdot\rho(\lambda)\cdot\Delta\lambda
\end{equation}
b) backward :
\begin{equation}
	dP_{ram,b} = dP_{ram}(\lambda,x)\cdot e^{-\alpha\cdot x}
\end{equation}
integrating over whole fibre 
\begin{equation}
	\Rightarrow P_{ram,b}= P_{in}\cdot e^{-\alpha\cdot x}\cdot\frac{sinh(\alpha\cdot L)}{\alpha}\cdot\rho(\lambda) \cdot\Delta\lambda
\end{equation}
In order to obtain the detection probabilities per gate ($p_{ram,f}$ and $p_{ram,b}$ respectively), used for the $QBER$ calculation (see Appendix~B), we calculate
\begin{equation}
	p_{ram,f}=\frac{P_{ram,f}}{h\nu}\cdot\eta\cdot\Delta t_{gate}
	\label{ramprobfw}
\end{equation}
where $\Delta t_{gate}$ is the gate duration and $\eta$ the detector efficiency. By replacing $P_{ram,f}$ by $P_{ram,b}$, one obtains $p_{ram,b}$ in the same manner.

\subsection{Explicit $QBER$ and key rate formulas}\label{sec:AppendixQBER}
The general definition of $QBER$ is given by the ratio between the number of false detections and total detections (right+false):
\begin{equation}
	QBER=\frac{false}{right+false}
\end{equation}
Introducing a sifting protocol specific parameter $\beta$ which is $\beta_{BB84}=1$ for BB84 and $\beta_{SARG}=\frac{2-V}{2}$ for SARG we obtain
\begin{equation}
	QBER = \frac{1}{2}\cdot\frac{p_\mu(1-V)+2\cdot p_{dc}+p_{AP}+p_{ram}+p_{ct}}{\beta\cdot p_\mu+2\cdot p_{dc}+p_{AP}+p_{ram}+p_{ct}}
\end{equation}
where every quantity signifies a detection probability per detector gate. In particular~: $p_\mu$ = signal detection, $p_{dc}$ = dark count, $p_{AP}$ = after pulse, $p_{ram}= p_{ram,f}+p_{ram,b}$ = Raman photon detection (see Eq.~\ref{ramprobfw}), $p_{ct}$ = cross talk photon detection and $V$ is the interference visibility. The signal detection probability $p_\mu$ is a product of the average number of photons per pulse $\mu$, fibre transmission $t$, detector efficiency $\eta$ and $t_{B}$ the loss of Bob's internal components. The optimal $\mu$ also depends on the sifting protocol, it is $\mu_{BB84}=t$ and $\mu_{SARG}=2\cdot\sqrt{t}$. 

We estimate the secret key rate after error correction and privacy amplification by
\begin{equation}
R_{sec} = R_{sift}\left(I_{AB}-I_{AE}\right).
\label{eq:DefSecretRateApp}
\end{equation}
Here, $R_{sift}$ is the sifted bit rate,
\begin{equation}
\label{eq:siftRate}
	R_{sift}=\frac{\frac{1}{2}\left(\beta\cdot p_{\mu} +2p_{dc}+p_{AP}+p_{ram}+p_{ct}\right)\cdot f_{rep}\cdot\eta_{duty}}{1+\tau_{dead}\left(p_{\mu}+2p_{dc}+p_{AP}+p_{ram}+p_{ct}\right)\cdot f_{rep}},
\end{equation}
where $f_{rep}$ is the pulse repetition frequency and $\eta_{duty}=\frac{L_{S}}{L+2L_{S}}$ accounts for the duty cycle of our system with $L$ the fibre length and $L_{S}$ the length of Alice's storage line. We note, that Eq.~\ref{eq:siftRate} does not account for double detections and Poissonian photon number statistics.

During error correction and privacy amplification, $R_{sift}$ is reduced by a factor $\left(I_{AB}-I_{AE}\right)$. $I_{AB}$ and $I_{AE}$ are the mutual information per bit between Alice and Bob, and between Alice and a potential eavesdropper, respectively.
Due to quantum bit errors, $I_{AB}$ is smaller than 1 and amounts to
\begin{eqnarray}
	\label{eq:IAB}
	I_{AB}=1-\eta_{ec}\cdot H\left(QBER\right),
\end{eqnarray}
with $H\left(p\right)=-p\log_{2}p-\left(1-p\right)\log_{2}\left(1-p\right)$ the binary entropy function. In the ideal case, the amount of bits discarded during error correction is given by the Shannon limit, i.e. $\eta_{ec}=1$. In practice, however, we observe that the implemented algorithm for CASCADE error correction consumes about 20~\% more bits than given by the Shannon limit. Hence, we correct Eq.~\ref{eq:IAB} by choosing $\eta_{ec}^{Cascade}=\frac{6}{5}$.

To calculate the information per bit $I_{AE}$ between Alice and an eavesdropper we assume that an eavesdropper has full control over the quantum channel (i.e. the visibility and fibre transmission). In contrast, he can not modify the characteristics of Bob's detectors. Additionally, we suppose that he performs an optimal coherent attack \cite{Fuchs1997} on pulses containing one photon, and a PNS attack \cite{PNS2000} if more than one photon is present in a pulse (without affecting the total detection rate at Bob). For BB84 with weak laser pulses one then obtains \cite{Niederberger2005}
\begin{eqnarray}
I_{AE,BB84}=\frac{\left(1-\frac{\mu}{2t}\right)\left(1-H\left(P\right)\right)+\frac{\mu}{2t}}{1+\frac{2p_{dc}}{\mu t \eta}}
\end{eqnarray}
with $P=\frac{1}{2}+\sqrt{D\left(1-D\right)}$, $D=\frac{1-V}{2-\mu / t}$. Based on the same assumptions, we use the results in \cite{BranciardSARG} to estimate for the SARG protocol
\begin{eqnarray}
I_{AE,SARG}=I_{pns}\left(1\right)+\frac{1}{12}\frac{\mu^{2}}{t}e^{-\mu}\left(1-I_{pns}\left(1\right)\right),
\end{eqnarray}
where $I_{pns}(k)=1-H(\frac{1}{2}+\frac{1}{2}\sqrt{1-\frac{1}{2^{k}}})$ is the potential information gain of an eavesdropper due to PNS attacks on multi-photon pulses when $k$ photons are split and stored.

\bibliography{Bibliography}
\end{document}